# Hypoxia-inducible factor 1α protects peripheral sensory neurons from diabetic peripheral neuropathy by suppressing accumulation of reactive oxygen species.

Daniel Rangel Rojas[1], Irmgard Tegeder[2], Rohini Kuner[1], Nitin Agarwal[1*]

[1]Institute of Pharmacology, Heidelberg University, Im Neuenheimer Feld 366, D-69120 Heidelberg, Germany.

[2]Institute for Clinical Pharmacology, Goethe-University Hospital, Frankfurt, Germany.

*Address correspondence to N.A, Institute of Pharmacology, Heidelberg University, Im Neuenheimer Feld 366, D-69120 Heidelberg, Germany. E mail: nitin.agarwal@pharma.uni-heidelberg.de

## ABSTRACT

Diabetic peripheral neuropathy (DPN) is one of the most common diabetic complications. Mechanisms underlying nerve damage and sensory loss following metabolic dysfunction remain large unclear. Recently, hyperglycemia-induced mitochondrial dysfunction and the generation of ROS have gained attention as possible mechanisms of organ damage in diabetes. Hypoxia-inducible factor 1 (HIF1α) is a key transcription factor activated by hypoxia, hyperglycemia, nitric oxide as well as ROS, suggesting a fundamental role in DPN susceptibility. Genetically-modified mutant mice, which conditionally lack HIF1α in peripheral sensory neurons (SNS-HIF1α$^{-/-}$), were analyzed longitudinally up to 6 months in the streptozotocin (STZ) model of type1 diabetes. Behavioral measurements of sensitivity to thermal and mechanical stimuli, quantitative morphological analyses of intraepidermal nerve fiber density and measurements of reactive oxygen species (ROS) in sensory neurons in vivo were undertaken over several months post-STZ injections to delineate the role of HIF1α

**Abbreviations:** HIF1α - Hypoxia-inducible factor 1α; ROS - reactive oxygen species; STZ - Streptozotocin; DPN - Diabetic peripheral neuropathy; DRG - dorsal root ganglia; SNS – sensory neuron-specific in DPN. Longitudinal behavioral and morphological analyses at 5, 13 and 24 weeks post-STZ treatment revealed that SNS-HIF1α$^{-/-}$ developed stronger hyperglycemia-evoked losses of peripheral nociceptive sensory axons associated with stronger losses of mechano- and heat sensation with a faster onset than HIF1α$^{fl/fl}$ mice. Mechanistically, these histomorphologic and behavioral differences were associated with significantly higher level of STZ-induced production of ROS in sensory neurons of SNS-HIF1α$^{-/-}$ mice as compared with HIF1α$^{fl/fl}$. Our results indicate that HIF1α is as an upstream modulator of ROS in peripheral sensory neurons and exerts a protective function in suppressing hyperglycemia-induced nerve damage by limiting ROS levels. HIF1α stabilization may be thus a new strategy target for limiting sensory loss, a debilitating late complication of diabetes.

## KEY MESSAGES

- Impaired Hypoxia-inducible factor 1α (HIF1α) signaling leads to early onset of STZ-induced loss of sensation in mice
- STZ-induced loss of sensation in HIF1α is associated with loss of sensory nerve fiber in skin.
- Activation of HIF1α signaling in diabetic mice protects the sensory neurons by limiting ROS formation generated due to mitochondrial dysfunction.

**Abbreviations:** HIF1α - Hypoxia-inducible factor 1α; ROS - reactive oxygen species; STZ - streptozotocin; DPN - Diabetic peripheral neuropathy; DRG - dorsal root ganglia; SNS – sensory neuron-specific

## INTRODUCTION

Diabetes is a chronic progressive metabolic disease characterized by elevated levels of blood glucose. It is one of the largest health care problems worldwide, with more than 422 million diabetic patients [1]. Diabetic peripheral neuropathy (DPN) is one of the most common diabetic complications. Patients with DPN experience pain, tingling, prickling sensations and numbness, which can progress to complete loss of sensation and motor dysfunction. These late diabetic complications in combination with vascular damage are the leading cause of lower extremity amputations [2, 3]. Clinical research from last two decades indicates that rigorous control of glucose reduces the incidence of DPN, suggesting a major role for hyperglycemia [4].

Uncovering molecular mechanisms underlying DPN remains a major challenge and a lack of mechanistic understanding has long hampered effective therapy of DPN. Only recently, hyperglycemia-induced mitochondrial dysfunction and the subsequent generation of ROS have gained attention as possible mechanisms [5]. Prolonged hyperglycemia energizes activation of the polyol [6], and hexosamine pathways [7], activation of PKC isoforms [8] and production of advanced glycation end products (AGE) [9, 10], thereby cumulatively increasing the intracellular ratio of NADH/NAD+. It has been hypothesized that this depletion of cellular antioxidant potential leads to cellular damage, which is marked by the progression of DPN [11]. Hyperglycemia-induced ROS formation causes the development of a pseudohypoxia state, which modulates intracellular signaling pathways such as mitogen activation protein kinases (MAPKs), nuclear factor kappa B (NF-κB), activator protein I (AP-I) and importantly, hypoxia-inducible factor-1 (HIF-1) [12, 13].

HIF1 is a heterodimeric transcriptional factor composed of α and β subunits (HIF1α and HIF1β respectively) [14, 15]. HIF1α is an oxygen-sensitive subunit and its activity is tightly regulated by cellular oxygen concentration [15], while HIF1β is constitutively expressed. In conditions of normoxia, HIF1α is subjected to oxygen-dependent hydroxylation at two prolyl residues, which enables its binding to Hippel-Lindau protein (VHL) leading to polyubiquitination and subsequent proteasomal degradation [16, 17]. In contrast, under hypoxic conditions, hydroxylation of HIF1α is inhibited, resulting in accumulation of HIF1α and its heteromerization with HIF1β. In consequence, stabilized transcription factor HIF1 binds to hypoxia response elements (HREs) and regulates the expression of target genes, leading to biological consequences.

In contrast to the clarity on hypoxia-mediated modulation of HIF1α, the role of hyperglycemia in the modulation of HIF1α is subject to much debate. Previous studies report contradictory observations, e.g. hyperglycemia-induced stabilization of HIF1α in pancreatic cancer cells versus its inhibition in primary dermal fibroblasts [33, 45]. Moreover, there is no knowledge on the role of HIF1α in hyperglycemia-induced dysfunction in peripheral sensory neurons in the course of DPN. Recently, it was reported that HIF1α deletion results in exaggerated acute thermal and cold nociception [18] in naïve mice, suggesting that HIF1α activity is important in peripheral nociceptive neurons. However, to date, the role of HIF1α in pathological DPN-related structural and functional changes in peripheral sensory neurons has not been studied.

Here, we investigated the role of HIF1α in peripheral pain-sensing (nociceptive) neurons over the chronic progression of DPN in mice genetically lacking HIF1α conditionally in the dorsal root ganglia (DRG) by undertaking detailed longitudinal analyses spanning 6 months following induction of type1 diabetes in the streptozotocin (STZ) model. Our results uncover a previously undescribed protective role of HIF1α in nerve damage and sensory loss associated with chronic stages of DPN and lead to the surprising observation that HIF1α is an upstream modulator of ROS levels and oxidative stress in peripheral sensory neurons under diabetic conditions.

## MATERIAL AND METHODS:

### Animal experiments

Age-matched 7-9 wks old SNS-HIF1α$^{-/-}$ and HIF1α$^{fl/fl}$ mice were used for experiments. Animals were maintained in a temperature controlled room with a light/dark cycle (12/12). Four to five littermates were housed in individually ventilated cage-rack system (Techniplast, Italy) with free access to water and food. All animal experiments were done in accordance with ethical guidelines and approved by Regierungspräsidium Karlsruhe, Germany.

## Sensory-neuron-specific knockout mice

Mice lacking HIF-1α (SNS-HIF1α$^{-/-}$) in a sensory neuron-specific manner were described previously [18]. Briefly, mice carrying homozygous alleles of HIf1α gene (HIF1α$^{fl/fl}$) were mated with mice expressing the Cre recombinase selectively in nociceptors (SNSCre) [19] to obtain homozygous SNS-HIF1α$^{-/-}$ and HIF1α$^{fl/fl}$ (control littermates). We have previously described the basic characterization of these mice in the study by Kanngiesser et al [18].

## Rodent model of type 1 diabetic model

Diabetes was induced by low dose multiple injections of STZ (60 mg/kg/d) in citrate buffer, intraperitoneally (i.p), in 8 wks-old SNS-HIF1α$^{-/-}$ or HIF1α$^{fl/fl}$ mice for 5 consecutive days. Blood glucose levels were measured once weekly using a glucometer (Accu-Chek, Roche Diagnostics) for the entire course of the experiment. Mice above > 350 mg/dl were considered to be diabetic. Mice were analyzed over a period of 5 wks to 24 wks post-STZ injection. At 2 weeks (wks) post-STZ injection, we achieved blood glucose between 380 and 500 mg/dl. The blood glucose levels were maintained in a range between 380 and 480 mg/dl by administering insulin sub-cutaneously as required over the entire course of an experiment to ensure uniformity and consistency.

## Behavioral measurements

All behavioral experiments were done in awake, unrestrained, acclimatized, age-matched adult mice. Mice were habituated to the experimental setup twice a day for 4 days prior to the behavioral testing. The experimenter was blinded for the genotype of the groups tested in the experiment. Thermal sensitivity was measured by recording paw withdrawal latency on application of infrared (IR) heat source on the plantar surface of hindpaw. A cut off of 20 sec was set to avoid burning of tissue. Two consecutive heat applications are performed at a time interval of 5 minutes. Von Frey measurements were done to assess mechanical sensitivity as described previously [20]. Briefly, mice were placed on an elevated grid and von Frey monofilaments exerting a specific force of 0.07, 0.16, 0.4, 0.6, 1.0, 1.4, 2.0 and 4.0 g were tested on the plantar hindpaw. Each monofilament was applied for 5 times at time interval of 5 min on the plantar hindpaw. 60% response frequency was as “thresholds” as described previously at basal, 14 and 24 wks post-STZ injection.

## ROS measurements in DRG

To detect production of ROS and superoxide in DRG of SNSHIF1α$^{-/-}$ and HIF1α$^{fl/fl}$ diabetic and nondiabetic mice, we intrathecally (i.t) injected MitoTrackerRedCM-$H_2$XROS dye (100 nm, 10 µl, Life Technologies). Twenty four h later, mice were perfused with 1x PBS and 4% PFA and DRG were extracted and 16 µm sections were made on cryotome [21]. The dye is in reduced form and non-fluorescent in the basal state and becomes fluorescent upon oxidation [22]. Presence of MitoTrackerRedCM-$H_2$XROS fluorescence at different point of time point’s post-STZ injection was analyzed using confocal microscopy.

## Immunohistochemistry

Immunohistochemistry was performed on punch biopsies of the plantar surface of the hindpaws in diabetic and non-diabetic HIF1α$^{fl/fl}$ an SNS-HIF1α$^{-/-}$ mice prior to and different time points post-STZ. The mice were perfused with 4% paraformaldehyde (PFA) and plantar skin was post-fixed in 4 % PFA for 24 hrs at 4°C. After overnight incubation in 30% sucrose, 16 µm cryo sections were made. The sections were permeabilized in 0.5% PBST, washed and blocked with 7% Horse serum (HS). The sections were incubated overnight with anti-CGRP (1:1000, sigma) primary antibody in 7% HS in PBS at 4°C [23]. Subsequently, sections were washed and incubated in Alexa Fluor-594-conjugated secondary antibody. Further, the sections were washed and mounted in Mowiol. Fluorescence images were obtained using a laser-scanning spectral confocal microscope and maximal projections were created using Leica SP8 software (Leica TCS SP8 AOBS, Bensheim, Germany). Same conditions were used for imaging of sections from control and test groups. The acquired images were analyzed using ImageJ software. The fluorescence intensity from the epidermal area was measured and the background staining from IgG control was deducted from each test sample.

### Statistical analysis

All the data were calculated and are presented as mean ± SEM. ANOVA for repeated measures followed by Bonferroni's test for multiple comparisons was employed to determine statistically significant differences. Changes with $p \leq 0.05$ were considered to be significant.

## RESULTS

### In vivo modulation of nociception by HIF1α in STZ model of type 1 diabetes

This study investigated the role of HIF1α in the modulation of levels of ROS in dorsal root ganglia in diabetic and non-diabetic mice. In order to address a potential contribution of HIF1α to DPN, we employed a mouse line with conditional genetic loss of HIF-1α selectively in a sensory neuron-specific manner (SNS-HIF1α$^{-/-}$ mice [18, 19]. In this line, we have previously demonstrated the deletion of HIF1α in peripheral nociceptive neurons via quantitative RT-PCR and *in situ* hybridization [18]. The STZ model of type 1 is characterized by slowly progressive β cell death and subsequently lymphocytic infiltration of pancreatic islets [26]. To investigate the role of HIF-1α in diabetes-induced ROS modulation, we employed a protocol of STZ injections, which gradually induces hyperglycemia and precludes acute neurotoxic effects that may result from high doses of STZ [24, 25]. Using this protocol, hyperglycemia developed to comparable levels in SNS-HIF1α$^{-/-}$ and HIF1α$^{fl/fl}$ mice (Fig 1a). Behavior measurements in STZ-injected mice are an important and direct measure of the progression of DPN. The STZ protocol reproduces sensory phenomena of human DPN, i.e. hyperalgesia, which is evident over 5-8 weeks post-STZ, and progressive loss of sensory functions, which is evident starting 20 wks post-STZ injection [27, 28]. We performed long-term studies in SNS-HIF1α$^{-/-}$ and corresponding control mice (HIF1α$^{fl/fl}$) in the STZ model to determine the course of behavioral changes, setting a focus on 5, 13 and 24 wks post-STZ injection. Thus, the time points we chose are temporally separated from potential acute toxic effects of STZ.

At 5 wks post-STZ, we observed that both SNS-HIF1α$^{-/-}$ and HIF1α$^{fl/fl}$ develop similar levels of hypersensitivity to mechanical stimuli applied to plantar surface of paw in form of graded strength von Frey hairs, evident as a decrease in mechanical nociceptive thresholds (figure 1b, c). Similarly, the development of thermal hyperalgesia, which is evident as a decrease in latency to an infrared heat lamp applied to the hindpaw plantar surface (figure 1b, c). In contrast, at 13 and 24 wks post-STZ injection, deletion of HIF1α in sensory neurons of the DRG had a marked impact on the development of sensory losses. Upon measuring mechanical response thresholds at 13 and 24 wks post-STZ, we observed that SNS-HIF1α$^{-/-}$ showed hypoalgesia as compared to HIF1α$^{fl/fl}$ mice, indicating an increase in the magnitude of STZ-induced sensory losses (figure 1b,*****, # p<0.0001, ANOVA, Bonferroni's test). Similarly, in the Hargreaves test, the time required for paw withdrawal upon stimulation with radiant heat was significantly increased in STZ-treated SNS-HIF1α$^{-/-}$ mice as compared to STZ-treated HIF1α$^{fl/fl}$ mice (figure 1c, ***** p<0.05, ******* p<0.0001, # p<0.01, ANOVA, Bonferroni's test). Hence, SNS-HIF1α$^{-/-}$ mice developed faster onset of hypoalgesia and thus showed a stronger loss of sensory function as compared to control mice, which is an indicative of an accelerated course of DPN upon loss of HIF1α specifically in peripheral sensory neurons.

## Manifestation of DPN in SNS-HIF1α$^{-/-}$ mice post-STZ injection

In order to complement behavioral abnormalities with actual changes in nerve morphology, we conducted a detailed neuropathological analysis of changes in epidermal nerve fiber density. We performed immunostaining of markers for nociceptive nerves, such as Calcitonin Gene-Related Peptide (CGRP), on skin sections in a longitudinal analysis up to 24 wks post-STZ. We analyzed DPN-associated reductions in the density of epidermal nociceptive innervation, measured by the fluorescence intensity of CGRP-positive fibers in epidermal region of the glaborous skin derived from the paw. The magnitude of this loss was higher and the progression faster in SNS-HIF1a$^{-/-}$ mice as compared to HIF1α$^{fl/fl}$ mice (Fig. 2a, b, ***** p<0.0001, # p<0.01, ANOVA, Bonferroni's test), at 13 and 24 wks post-STZ injection, indicating an aggravation of DPN manifestations in the absence of HIF1α.

## Modulation of ROS levels in DRGs of diabetic mice by HIF1α

Previous studies have established that hyperglycemia modulates the respiratory chain of the mitochondria and augments oxidative stress [29]. Overproduction of ROS has been shown to modulate HIF1 function [30], with both positive and negative modulation being reported, and several mechanisms have been proposed [31, 32, 16]. Despite these insights on the regulation of HIF1 by ROS, none of the studies investigated the impact of HIF1 function on modulation of ROS levels itself. Therefore, we measured the levels of ROS in vivo in DRG neurons of SNS-HIF1α$^{-/-}$ and control HIF1α$^{fl/fl}$ mice in the basal state (i.e. naïve mice) and at 13 or 24 wks post-STZ treatment. We intrathecally injected MitoTrackerRedCM-$H_2$XROS (100 nm, 10 µl), a reporter for ROS generation [22], and observed that STZ-induced DPN was linked to a significant increase in MitoTrackerRedCM-$H_2$XROS fluorescence in DRG neurons, suggesting ROS and superoxide build-up (Fig. 3a, b, *, # p<0.0001, ANOVA, Bonferroni's test). Importantly, the extent of ROS and superoxide accumulation was significantly increased in SNS-HIF1α$^{-/-}$ mice as compared to HIF1α$^{fl/fl}$ mice at 13 and 24 wks of STZ injection (Fig. 3a, b). Thus, a selective deletion of HIF1α resulted in inflated levels of ROS formation in peripheral sensory neurons in vivo at time points, which temporally matched the exacerbation of DPN in SNS-HIF1α$^{-/-}$ mice.

## DISCUSSION

The main insights revealed by this study are: (i) the transcription factor HIF1α plays a protective role against axonal pathology and loss of peripheral nerve fibers in DPN, (ii) a loss of HIF1α in sensory neurons of diabetic mice leads to a faster progression of DPN-associated loss of sensation at the level of organismal behavior; (iii) a loss of HIF1α results in an enhanced build-up of ROS in sensory neurons, thereby placing ROS downstream of HIF1α.

HIF1α has only recently been suggested to modulate pain following direct mechanical injury to peripheral nerves [18]. However, to date, nothing was known about the role of HIF1α in pain and nerve damage caused by metabolic dysfunction. Indeed, there is a large mechanistic diversity between diverse forms of pathological pain, such as that arising from direct mechanical injury to nerves that is focal and localized versus metabolically-induced nerve dysfunction in the course of diabetes, which is widespread and diffuse. Very little is known about how nerve damage develops under metabolic stress. We hypothesized that HIF1α may be a vital link because it has been reported to be directly regulated by hyperglycemia in some tissues, such as endothelial cells and cardiac myocytes, with both positive and negative modulation being reported [33]. Indeed, using highly selective genetic perturbations, we observed that both structural remodeling of nerves and their functionality (via behavioral analysis) were highly affected upon a loss of HIF1α in sensory neurons. Interestingly, only chronic changes, such as structural damage and the ensuing sensory loss at about 4-6 months post-diabetes induction were affected in SNS-HIF1α$^{-/-}$ mice, emphasizing the strength of longitudinal analyses. Particularly this late component of diabetic changes is key in the clinical situation, frequently resulting in small and large fiber neuropathies, diabetic feet and amputations [34]. In contrast, early STZ-evoked hypersensitivity was not changed upon a loss of HIF1α in sensory neurons. Several recent studies have addressed mechanisms of early hyperalgesia in diabetic models and there is converging evidence that this involves acute modulation of ion-channels, such as members of the Transient Receptor Potential (TRP) family, such as TRPV1 and TRPA1 by oxidative or lipid metabolites [35, 36].

The nature of modulation of DPN by HIF1α is of much interest. In terms of injury induced by hypoxia and/or hyperglycemia, both protective and destructive roles have been observed for HIF1α depending upon the type of tissue and the context [33, 37]. Here, we observed a protective function for HIF1α against nerve damage induced by metabolic dysfunction in a model of type 1 diabetes. These observations were further strengthened by measurement of ROS levels in DRG, which revealed that the deletion of HIF1α results in increased accumulation of ROS in sensory neurons as compared to controls over chronic phases of diabetes. This is unexpected, because previously published studies in other tissues reported the modulation of HIF1α activity by ROS, but there is little evidence for modulation of ROS levels by HIF1α. In kidney tissue under hyperglycemia conditions, ROS has been shown to impair HIF1α activation as well as expression via inhibition of NO and Rac1 [33]. Moreover, ROS can further inhibit HIF1α by increasing ubiquitin-proteasome activity [38]. In contrast, Bonello *et. al*. demonstrate that ROS leads to HIFα expression and activation in a NFκB-dependent manner in pulmonary artery smooth muscle cells [39]. In sensory neurons of chronically hyperglycemic mice, we observed that ROS is downstream of HIFα, which suggests that HIFα exerts its neuroprotective functions in this context by suppressing ROS. The precise mechanisms how HIFα affects ROS levels are not clarified in this study. Contributing mechanisms may include the role of HIF1α in modulating the expression of metabolic enzymes [40], which may decrease the metabolic overload in sensory nerves and suppress the generation of harmful ROS and free radical species.

Recent studies have implicated ROS in the modulation of acute pain [41] and the administration of ROS scavengers peripherally or spinally has antinociceptive effects in models of inflammatory and neuropathic pain [42]. It is believed that ROS production via mitochondrial respiratory chain is a causal link between hyperglycemia and pathways involved in hyperglycemia-induced tissue damage. Under hyperglycemic conditions, the increased electron flux in the respiratory chain leads to an increase in the ATP/ADP ratio and in the mitochondrial membrane potential, which result in partial inhibition of complex III and thereby induce an accumulation of electrons to generate potentially harmful complex-I ROS [29, 43]. It is postulated that accumulation of ROS may result in a bioenergetic failure, depletion of antioxidant defenses and neuroinflammation [44]. These mechanisms likely contribute to nerve damage and the ensuing sensory loss observed upon the excessive accumulation of ROS in diabetic mice lacking HIF1α.

## CONCLUSIONS

In summary, this study identifies HIF1α as an important protective molecule against DPN-associated long-term nerve damage and sensory loss. Moreover, we propose that HIF1α functions as an upstream suppressor of ROS levels in sensory neurons. These observations have direct relevance to a highly debilitating and intractable long-term complication of diabetes and hold significance for novel therapeutic measures designed to strengthen the protective role of HIF1α in sensory neurons over chronic diabetes.

## AUTHORS CONTRIBUTIONS

NA and RK designed the study; IT mated HIF1α$^{fl/fl}$ mice and SNS-Cre mice and gave general conceptual input; DRR performed and analyzed in vivo experiments; NA and DRR performed and analyzed in vitro experiments; NA and RK generated the SNS-Cre line and wrote the manuscript. All authors approved the final version.

## ACKNOWLEDGMENTS

The authors thank Rose LeFaucheur for secretarial help, Dunja Baumgartl-Ahlert and Hans-Joseph Wrede for technical assistance and Dr. Manuela Simonetti for help with i.t injections. This work was supported by a grant from the Deutsche Forschungsgemeinschaft (DFG) in the Collaborative Research Center 1118 (SFB1118 Project B06) to N.A. and R.K. We acknowledge support from the Interdisciplinary Neurobehavioral Core (INBC) for behavioral experiments.

## COMPLIANCE WITH ETHICAL STANDARDS

All animal experiments were done in accordance with ethical guidelines and approved by Regierungspräsidium Karlsruhe, Germany. The manuscript does not contain clinical studies or patient data.

## CONFLICT OF INTEREST

The authors declare that there is no conflict of interest.

## Figure legends:

**Figure 1:** Analysis of hyperglycemia and sensory dysfunction in mice lacking HIF1α selectively in peripheral nociceptive neurons **(**SNS-HIF1α$^{-/-}$) in the Streptozotocin (STZ) model of type 1 diabetes in a longitudinal study spanning 6 months. **(A)** Blood glucose levels in SNS-HIF1α$^{-/-}$ and HIF1α$^{fl/fl}$ mice after STZ injection **(B & C)** Course of DPN-associated sensory changes upon loss of HIF1α in peripheral nociceptive neurons. **(B)** Summary of response thresholds to mechanical stimuli (defined as a force eliciting a response of paw withdrawal at least 40% response rate) and **(C)** thermal latency before and at 5, 13 and 24 days after STZ treatment (60 mg/kg for 5 days) in SNS-HIF1α$^{-/-}$ or HIF1α$^{fl/fl}$ (n = 10 mice per group). Data are represented as mean ± S.E.M. *$p < 0.05$ compared to baseline, # $p < 0.05$

as compared to control group two-way ANOVA for repeated measurements with Bonferroni multiple comparison test.

**Figure 2**: Course and quantitative morphological analysis of DPN-associated pathology of peripheral nociceptive nerves in SNS-HIF1α$^{-/-}$ mice and control littermates. **(A)** Identification of intra-epidermal nerve fibers via immunostaining for CGRP in the paw skin of SNS-HIF1α$^{-/-}$ and HIF1α$^{fl/fl}$ mice after STZ injection. The dotted white lines represent the epidermal layer in the skin. **(B)** Quantification of reduction in epidermal nerve fiber density post-STZ injection in SNS-HIF1α$^{-/-}$ and HIF1α$^{fl/fl}$ mice. n=6 - 8 mice per group. Data are represented as mean ± S.E.M. *$p < 0.05$ compared to baseline, # $p < 0.05$ as compared to control group two-way ANOVA for repeated measurements with Bonferroni multiple comparison test. Scale bar represents 30 µm**.**

**Figure 3:** Quantitative analysis of ROS measurement in vivo in peripheral sensory neurons at stages corresponding to chronic nerve damage and sensory dysfunction in diabetic mice. (**A**) Examples and **(B)** quantitative analysis of MitoTrackerRed staining in DRGs of STZ-injected SNS-HIF1α$^{-/-}$ and HIF1α$^{fl/fl}$ mice after intrathecal delivery of the ROS reporter dye. Arrows indicate ROS-positive neurons; n=6 DRGs per group. *$p < 0.05$ compared to baseline, # $p < 0.05$ as compared to control group two-way ANOVA for repeated measurements with Bonferroni multiple comparison test. Scale bar represents 30 µm.

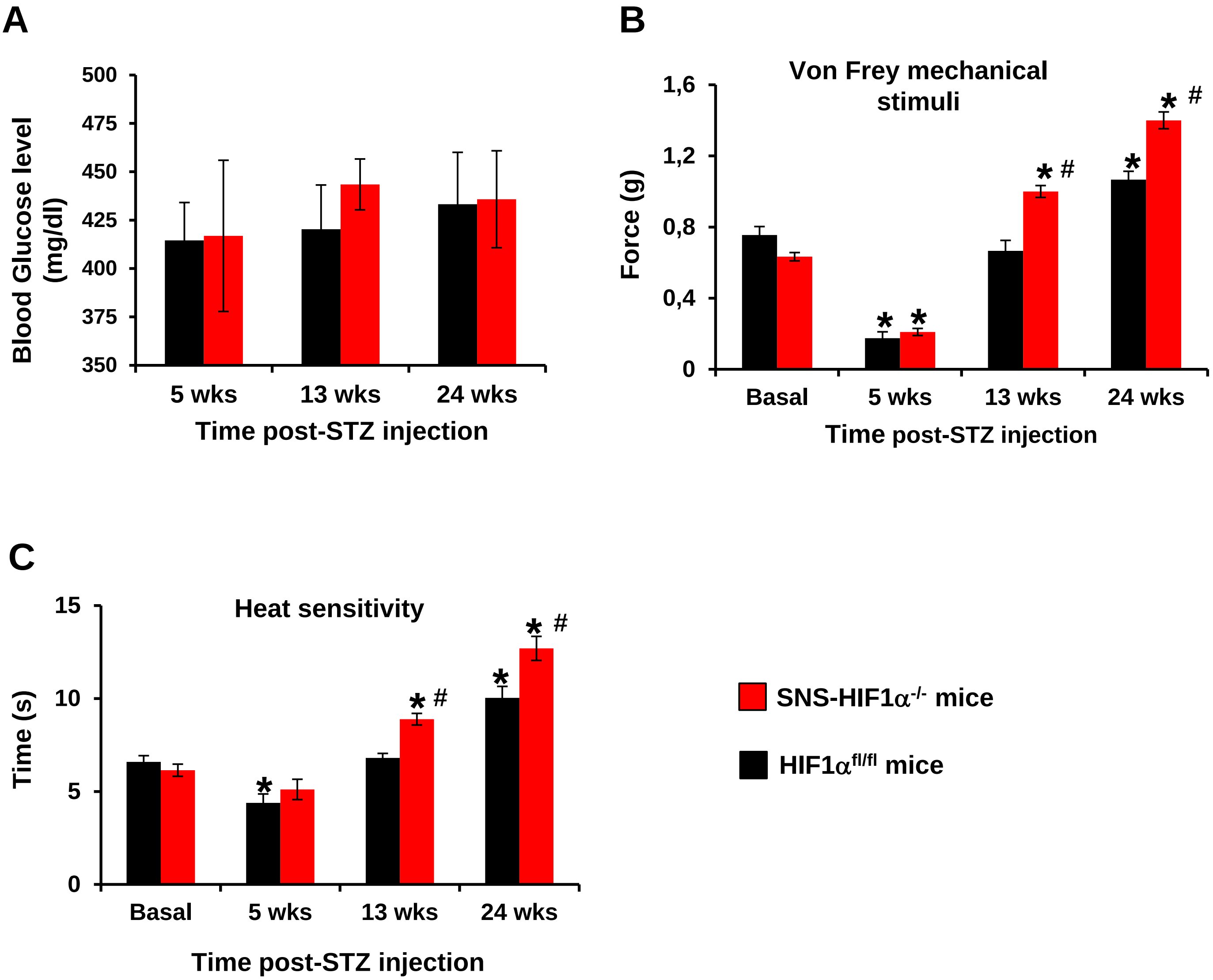

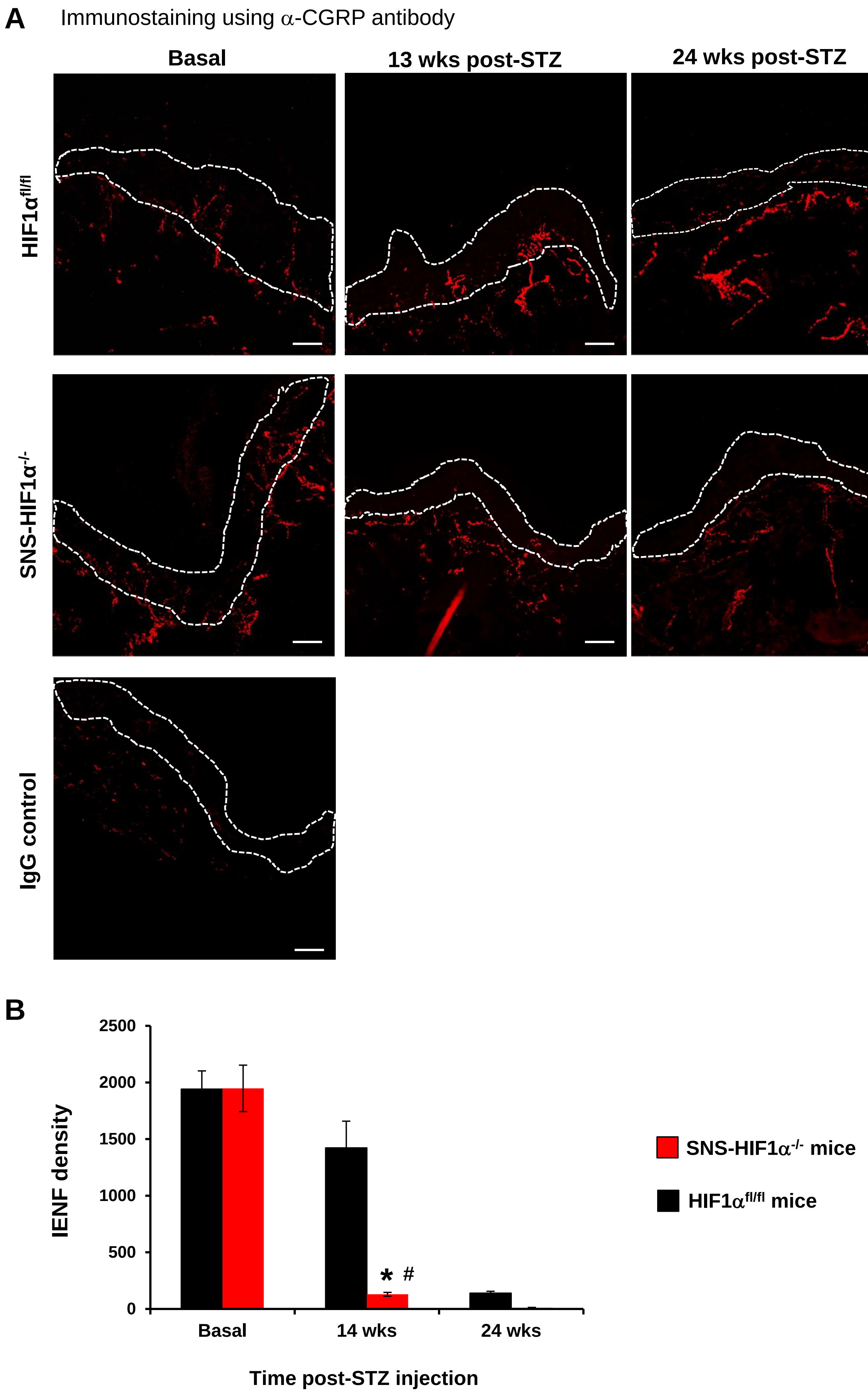

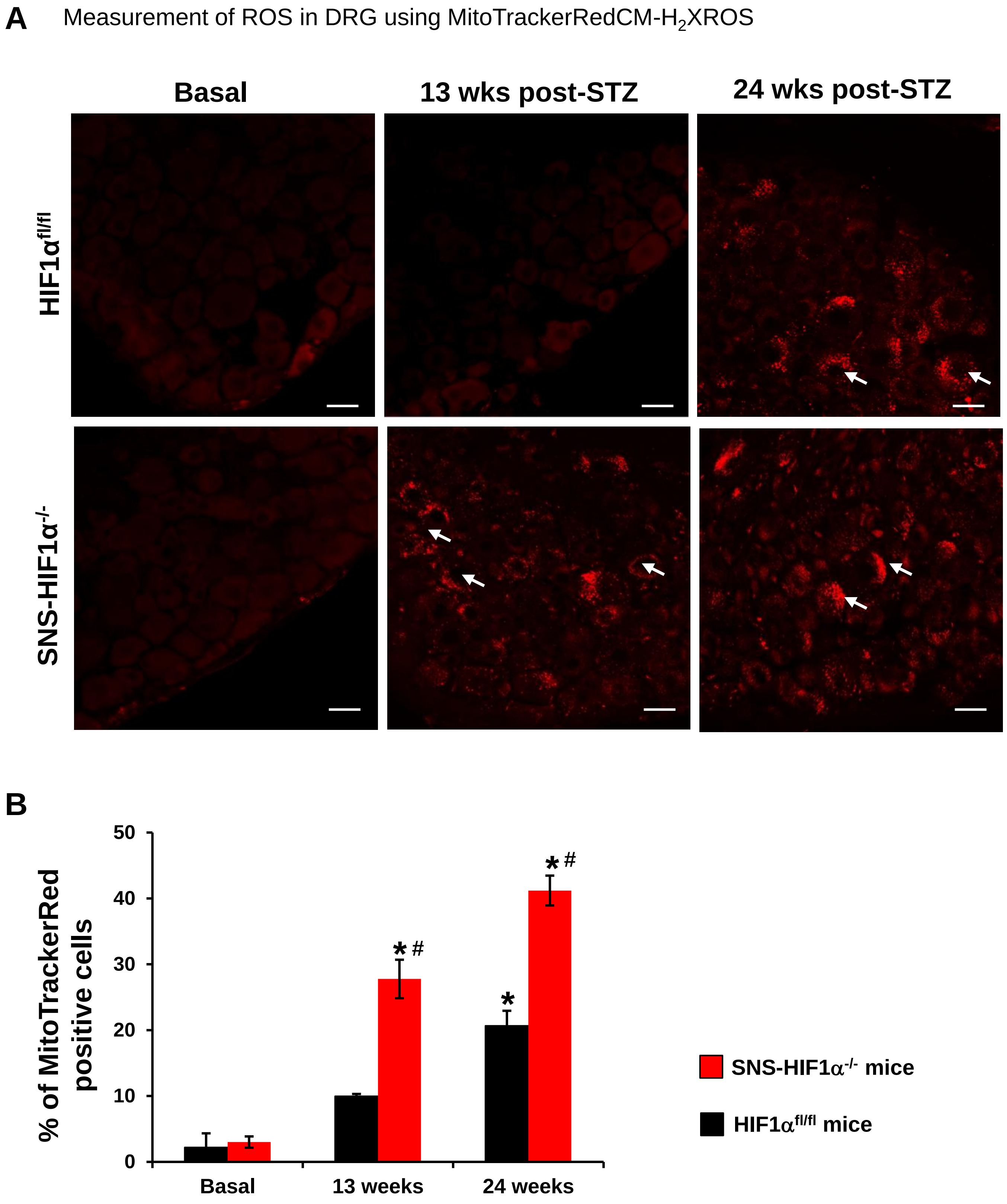